# Dual-Band Negative Index Metamaterial: Double-Negative at 813 nm and Single-Negative at 772 nm

U. K. Chettiar, A. V. Kildishev, H.-K. Yuan, W. Cai, S. Xiao, V. P. Drachev, and V. M. Shalaev

*Birck Nanotechnology Center, School of Electrical and Computer Engineering, Purdue University, West Lafayette, IN 47907, USA*

This work is concerned with the experimental demonstration of a dual-band negative index metamaterial. The sample is double-negative (showing both a negative effective permeability and a negative effective permittivity) for wavelengths between 799 and 818 nm of linearly polarized light with a real part of refractive index of about −1.0 at 813 nm; the ratio −Re(n)/Im(n) is close to 1.3 at that wavelength. For an orthogonal polarization, the same sample also exhibits a negative refractive index in the visible (at 772 nm). The spectroscopic measurements of the material are in good agreement with the results obtained from a finite element electromagnetic solver for the actual geometry of the fabricated sample at both polarizations.

The refractive index ($n = n' + \iota n''$) is the key parameter in the interaction of light with matter. Normally $n'$ has been considered to be positive, but the condition $n' < 0$ does not violate any fundamental physical law. Negative index metamaterials (NIMs) with $n' < 0$ have some remarkable properties, which are critical for a number of potential applications. Typically, a NIM is an artificially engineered metal-dielectric composite that exhibits $n' < 0$ within a certain range of wavelengths. A magnetic response should be observed in the NIMs within this range, since this is required to make the real part of the effective refractive index negative. It can be accomplished either through the strong (sufficient) condition that $\mu' < 0$ and $\varepsilon' < 0$, or through a more general (necessary) condition $\varepsilon'\mu'' + \mu'\varepsilon'' < 0$. The latter condition strictly implies that there cannot be $n' < 0$ in a passive metamaterial with $\mu = 1 + 0\iota$.

As follows from the above conditions, two types of NIMs can be introduced. A double negative NIM (DN-NIM), is a material with simultaneously negative real parts of its effective permeability and permittivity ($\varepsilon' < 0$ and $\mu' < 0$). A single negative NIM (SN-NIM) has a negative refractive index with either $\varepsilon'$ or $\mu'$ (but not both) being negative. In all the optical SN-NIM known to date, the real part of the permittivity is negative, whereas the real part of permeability is positive ($\varepsilon' < 0$ and $\mu' > 0$). At optical wavelengths getting $\varepsilon' < 0$ is easy compared with $\mu' < 0$, due to the fact that noble metals naturally have a negative permittivity above the plasma wavelength.

The ratio $-n'/n''$ is often taken as a figure of merit (FOM) for optimizing a NIM performance, since low-loss NIMs are desired. The FOM can be rewritten as $-(|\mu|\varepsilon' + |\varepsilon|\mu')/(|\mu|\varepsilon'' + |\varepsilon|\mu'')$, indicating that a DN-NIM with $\mu' < 0$ and $\varepsilon' < 0$ is better than a SN-NIM with the same $n' < 0$, but with $\mu' > 0$.

Substantial progress has been achieved recently in the field of NIMs. The first experimental demonstration was given in 2001 at microwave frequencies.[1] The first NIMs in the optical range were demonstrated in 2005 at 1.5 μm (Ref. [2]) and 2 μm.[3] In both cases it was a SN-NIM with a FOM of about 0.1 and 0.5 respectively. These results have been improved substantially in 2006; the first DN-NIM in the optical range was demonstrated at 1.4 μm with a FOM of about three[4] and at 1.8 μm with a FOM above 1.[5] Most recently, the negative refractive index was pushed into the visible with the demonstration of a negative index at 780 nm.[6] This was a SN-NIM with a maximum FOM of about 0.5. In this paper we present the results for a DN-NIM at the shortest wavelength to date. The DN-NIM has a maximum FOM of 1.3 at a wavelength of 813 nm. In addition, the same structure exhibits SN-NIM behavior at 772 nm, thus showing negative $n'$ at today's lowest visible wavelengths.

It has been previously shown[3,7,8] that a sub-wavelength bi-periodic cross-grating, consisting of two perforated thin metal layers separated by a thin dielectric, is a good design prototype for NIMs. This design was used for the initial optimization. The sample was fabricated using E-beam lithography followed by E-beam evaporation and lift-off. The structure was fabricated on ITO (Indium tin oxide) coated glass substrate. The ITO layer was 15 nm thick. Fig. 1a depicts a FE-SEM image of our structure, which is made of two 33-nm layers of perforated silver separated by a 38-nm layer of alumina. The lattice constant of the structure is 300 nm in both lateral directions. A 10 nm thick layer of alumina was deposited on top and below the structure to protect the silver layer from deterioration and improve adhesion to the substrate.

We first consider the double-negative regime of the structure. Fig. 1a depicts the "primary" polarization of incident light capable of creating the double-negative regime. The incident magnetic field is polarized along the set of wider parallel strips. These 'magnetic strips' are schematically shown in Fig. 1a as darker strips on top of the FE-SEM background image. The magnetic strips of the upper and lower layer are coupled at the magnetic resonance; in the optimized design the magnetic resonance should be sufficiently strong resulting in negative values for effective permeability. The other set of parallel strips is aligned with the incident electric field and shows no diffraction (since the period is subwavelength). The 'electric strips' are also schematically shown as lighter strips in Fig. 1a. Similar to the approach discussed in Ref.

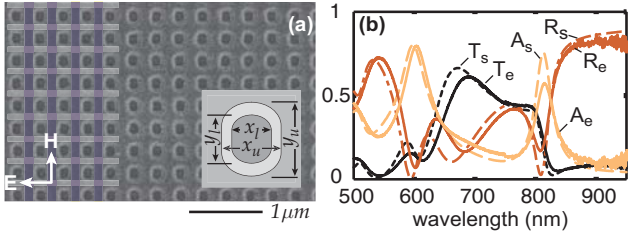

Fig. 1. (a) Background shows FE-SEM image of the NIM sample. The inset in Fig. 1a depicts the unit cell of the simulated structure. (b) A comparison of the experimental transmission ($T_e$), reflection ($R_e$) and absorbtion ($A_e$) spectra of the sample with simulated results ($T_s$, $R_s$, and $A_s$) at the primary linear polarization as shown in Fig. 1a.

[9], the electric strips behave like a "diluted metal", providing a controllable broadband negative permittivity. As a result, the structure can exhibit simultaneous negative permeability and permittivity, demonstrating a DN-NIM behavior.

To validate the design, the sample was optically characterized by obtaining its transmittance and reflectance spectra using normally incident light separately at both the primary and secondary linear polarizations. An ultra-stable tungsten lamp (B&W TEK BPS100) was used as a light source. The spectral range of the lamp covers the entire visible and near-infrared optical bands. A Glan Taylor prism was placed at the output of the broadband lamp to select light with a given linear polarization. The signal transmitted (or reflected) from the sample was introduced into a spectrograph (Acton SpectraPro 300i) and finally collected by a liquid nitrogen cooled CCD-array detector. The data were normalized to a bare substrate (for the transmission spectrum) and a calibrated silver mirror (for the reflection spectrum). The measured spectra for the DN-NIM regime (primary polarization) are shown in Fig. 1b. The figure compares optical measurements with the simulated data obtained from commercial finite element 3D full-wave electromagnetic solver (COMSOL MULTIPHYSICS) using high-order elements. The inset in Fig. 1a shows the unit cell of the simulated structure, where the stadium-shape void ($x_l = 122$ nm, $x_u = 218$ nm, $y_l = 154$ nm, and $y_u = 256$ nm) follows an averaged geometry obtained in fabrication. In simulations the permittivity of silver is taken from experimental data,[10] with the exception that $\varepsilon''$ is assumed to be three times that of bulk silver. This extra damping corresponds to the additional losses in the silver due to imperfections in metallic elements and metal-dielectric interfaces. This loss correction is done to match the experimental spectra; the averaged loss-adjustment factor of 3 has been successfully validated experimentally in recent studies.[3,6,11]

The experimental and simulated spectra shown in Fig. 1b demonstrate a good match over a broad range of measured wavelengths (from 500 to 950 nm) and include sharp resonant features: an electric resonance at about 600 nm and a magnetic resonance close to 800 nm. Beyond 950 nm signal detection is difficult due to a much lower signal-to-noise ratio.

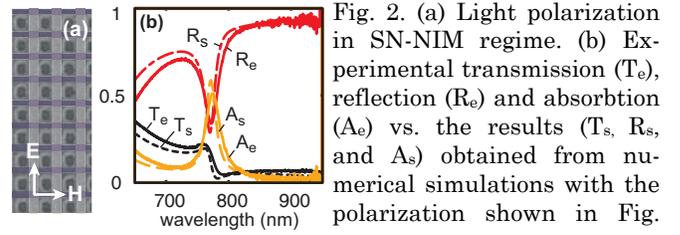

Fig. 2. (a) Light polarization in SN-NIM regime. (b) Experimental transmission ($T_e$), reflection ($R_e$) and absorbtion ($A_e$) vs. the results ($T_s$, $R_s$, and $A_s$) obtained from numerical simulations with the polarization shown in Fig.

Fig. 2a depicts the "secondary" linear polarization, which is orthogonal to the primary polarization of the incident light. In this case, the sample changes its behavior from double-negative to single-negative. In contrast to the DN-NIM regime shown in Fig. 1a, at the secondary polarization the incident electric field is aligned along the set of wider parallel strips. These 'electric strips' are schematically indicated in Fig. 2a as lighter strips over the FE-SEM image. The other set of parallel strips is aligned with the incident magnetic field. The 'magnetic strips' are also schematically shown in Fig. 2a as darker strips. Note that in the secondary polarization a different set of strips of the upper and lower layer is coupled at the magnetic resonance.

Experimental and simulated spectra for the secondary polarization also demonstrate a good match. A selected range of prime interest (around the magnetic resonance close to 770 nm) is shown in Fig. 2b. The unit cell and all optical constants used in the numerical simulations at the secondary polarization are identical to those used at the primary polarization. In comparison to the DN-NIM regime, the width of the magnetic strips in the single-negative case is smaller and the magnetic resonance is blue-shifted to 770 nm. In addition, the resonance is not sufficiently strong; only positive values for the effective permeability are obtained. This is confirmed by a much higher reflectance (and lower transmittance) before the resonance peak in absorbance indicating that the structure is far off from the impedance-matching condition. At this polarization the structure can exhibit only SN-NIM behavior with a positive permeability, in addition to the fact that the real part of the effective permittivity becomes even more metallic due to the wider 'electric' strips. Furthermore, the structure can demonstrate dual-band magnetic responses at a linear polarization of 45 degrees.

Figures 1b and 2b demonstrate good agreement between our spectroscopic measurements and numerical simulations obtained in a wide wavelength range for two different linear polarizations. This is a good indication of the validity of the numerical model, which is used to retrieve the effective parameters of the equivalent homogenized layer. The effective parameters are obtained by utilizing an extension of a standard homogenization technique[12] shown in Ref. [8] for a layer on a glass substrate. In this approach the thickness of the equivalent homogenized layer is equal to the physical thickness of the structure including the ITO layer ($\Delta = 139$ nm). The technique uses the complex values of reflection and transmission coefficients obtained either from the simulations or meas-urements.

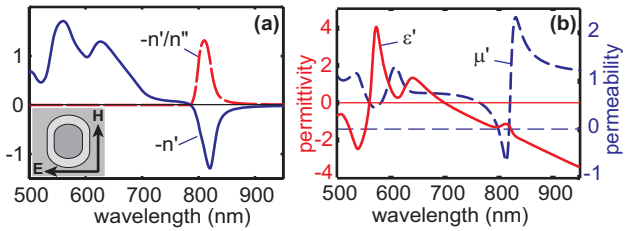

Fig. 3. The sample in DN-NIM regime. The primary polarization of light is used in modeling as shown in the inset. (a) The real part of the effective refractive index and FOM. (b) The real part of the effective permeability (μ') and permittivity (ε') are both negative from 799 to 818 nm.

The retrieved results for the primary polarization are shown in Fig. 3. The real part of the refractive index ($n'$) and the figure of merit ($-n'/n''$) are depicted in Fig. 3a (FOM is set to zero when $n' > 0$). The best FOM of 1.3 is obtained at a wavelength of 813 nm, where $n'$ is about –1.0. The minimal value of the refractive index, $n' \approx -1.3$, is achieved at 820 nm, but with a lower FOM of 0.9. As indicated in Fig. 3b, $\mu'$ is negative between 799 and 818 nm. This band is the DN-NIM regime. In this range $\varepsilon'$ deviates within –1.2±0.1. The strongest magnetic response ($\mu' \approx -0.7$) is obtained at a wavelength of 813 nm, where $\varepsilon' \approx -1.1$.

The magnetic resonance in periodic NIMs introduces an electric anti-resonant response close to the resonant wavelength.[8,9] Therefore, a reversed effect should be also observed within the electric resonance band, where the magnetic anti-resonance is present. In our sample this periodicity effect can be clearly observed in Fig. 3b, where an anti-resonance of $\varepsilon'$ coincides with the magnetic resonance around 820 nm.

The results for the SN-NIM regime obtained at the secondary polarization are shown in Fig. 4, where Fig. 4a depicts $n'$ and the ratio $-n'/n''$, while Fig. 4b shows $\mu'$ and $\varepsilon'$. At this polarization the best FOM of 0.7 is obtained at a wavelength of 772 nm, where $n'$ is about –0.9. The lowest value of the refractive index ($n' \approx -1.0$) is achieved at 776 nm, but with a slightly lower FOM. Fig. 4b shows that $\mu'$ never becomes negative. Its minimal value ($\mu' \approx 0.2$) is obtained at a wavelength of 769 nm along with $\varepsilon' \approx -2.0$. In comparison to the relatively narrow band (about 20 nm) of DN-NIM behavior, the wavelength range of the SN-NIM regime at the secondary polarization is much wider. This NIM band starts at about 753 nm and ends at about 810 nm with the value of $\varepsilon'$ decaying towards longer wavelengths from about –2.0 to –3.0. Similar to the DN-NIM case, the typical periodicity effect is also evident in Fig. 4b through the anti-resonant behavior of $\varepsilon'$ at the magnetic resonance around 780 nm.

In summary, a dual-band NIM sample with a period of 300 nm (in both directions) was fabricated and characterized using optical measurements. Depending on the polarization of normally incident light, different spectral behavior is demonstrated experimentally. Numerical simulations of the structure are in good agreement with optical measurements in a wide range of wavelengths for two orthogonal linear polarizations. At the primary polarization, the double-negative effective properties ($\varepsilon' < 0$ and $\mu' < 0$) are achieved in a 20-nm wavelength band around 810 nm; the lowest effective $\mu'$ of about −0.7 is observed at 813 nm along with an $n' \approx -1.0$ and a FOM of about 1.3. A second wavelength band of single-negative operation is found at the orthogonal "secondary" polarization, where the sample exhibits NIM behavior in the visible (772 nm) with $n' \approx -0.9$ and a FOM of about 0.7.

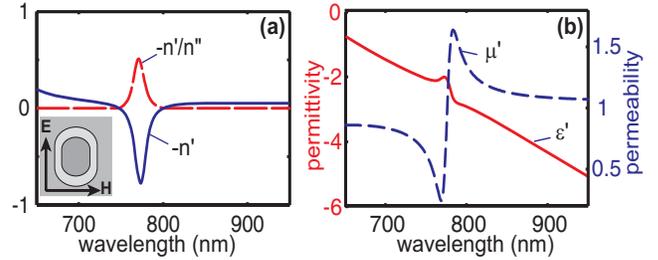

Fig. 4. The sample in SN-NIM regime. The secondary polarization of light is used in modeling as shown in the inset. (a) The real part of the effective refractive index and FOM. (b) The effective permeability (μ') is positive everywhere and only the permittivity (ε') is negative.

This work was supported in part by ARO grant W911NF-04-1-0350 and by ARO-MURI award 50342-PH-MUR.